\documentclass[a4paper,11pt]{article}
\usepackage{jheppubMOD} 
                     
\usepackage{lipsum}
\usepackage{graphicx}  
\usepackage{multirow}
\usepackage[x11names,dvipsnames]{xcolor}
\usepackage{orcidlink}

\hypersetup{
    colorlinks,
    linkcolor={jpac-blue},
    citecolor={jpac-blue},
    urlcolor={jpac-blue}
}

\usepackage{lmodern}
\usepackage{amsmath,amssymb}
\usepackage{mathtools}
\usepackage[numbers,sort&compress]{natbib}

\usepackage{tikz}
\usepackage{tikz-3dplot}
\usepackage{pgfplots,pgfplotstable}
\usetikzlibrary{pgfplots.groupplots}
\usetikzlibrary{decorations.markings}
\usetikzlibrary{shapes}
\usetikzlibrary{shapes.misc}
\usetikzlibrary{patterns}
\usepgfplotslibrary{fillbetween}
\pgfplotsset{compat=newest}
\usetikzlibrary{positioning}
\usetikzlibrary{calc}

\usepackage{lipsum}
\usepackage{stackrel}

\usepackage[compat=1.1.0]{tikz-feynman}

\usepackage{cleveref}
\usepackage{ifthen}
\usepackage{xspace}
\usepackage[shortlabels]{enumitem}

\usepackage[sicmds,freestanding]{hepunits}
\DeclareSIUnit{\fm}{\femto\metre}

\definecolor{jpac-blue}{RGB}{ 31,119,180}
\definecolor{jpac-red}{RGB}{214,39, 40}
\definecolor{jpac-green}{RGB}{ 44,160, 44}
\definecolor{jpac-orange}{RGB}{255,127, 14}
\definecolor{jpac-purple}{RGB}{148,103,189}
\definecolor{jpac-brown}{RGB}{140, 86, 75}
\definecolor{jpac-pink}{RGB}{227,119,194}
\definecolor{jpac-gold}{RGB}{188,189, 34}
\definecolor{jpac-aqua}{RGB}{ 23,190,207}
\definecolor{jpac-grey}{RGB}{127,127,127}

\newcommand\bsub{\begin{subequations}}
\newcommand\esub{\end{subequations}}

\renewcommand{\Re}{\text{Re}}
\renewcommand{\Im}{\text{Im}}

\newcommand{\kev}{\ensuremath{{\mathrm{\,ke\kern -0.1em V}}}\xspace}
\newcommand{\mev}{\ensuremath{{\mathrm{\,Me\kern -0.1em V}}}\xspace}
\newcommand{\gev}{\ensuremath{{\mathrm{\,Ge\kern -0.1em V}}}\xspace}
\newcommand{\gevsq}{\ensuremath{{\mathrm{\,Ge\kern -0.1em V}^2}}\xspace}
\newcommand{\tev}{\ensuremath{{\mathrm{\,Te\kern -0.1em V}}}\xspace}

\def\XXint#1#2#3{{\setbox0=\hbox{$#1{#2#3}{\int}$}\vcenter{\hbox{$#2#3$}}\kern-.5\wd0}}

\newcommand{\moda}{\left\lvert a \right\rvert}

\begin{document}

\title{\boldmath $\phi \to 3\pi$ and $\phi\pi^{0}$ transition form factor from Khuri-Treiman equations}

\collaboration{JPAC Collaboration}
\collaborationImg{\includegraphics[height=2cm,keepaspectratio]{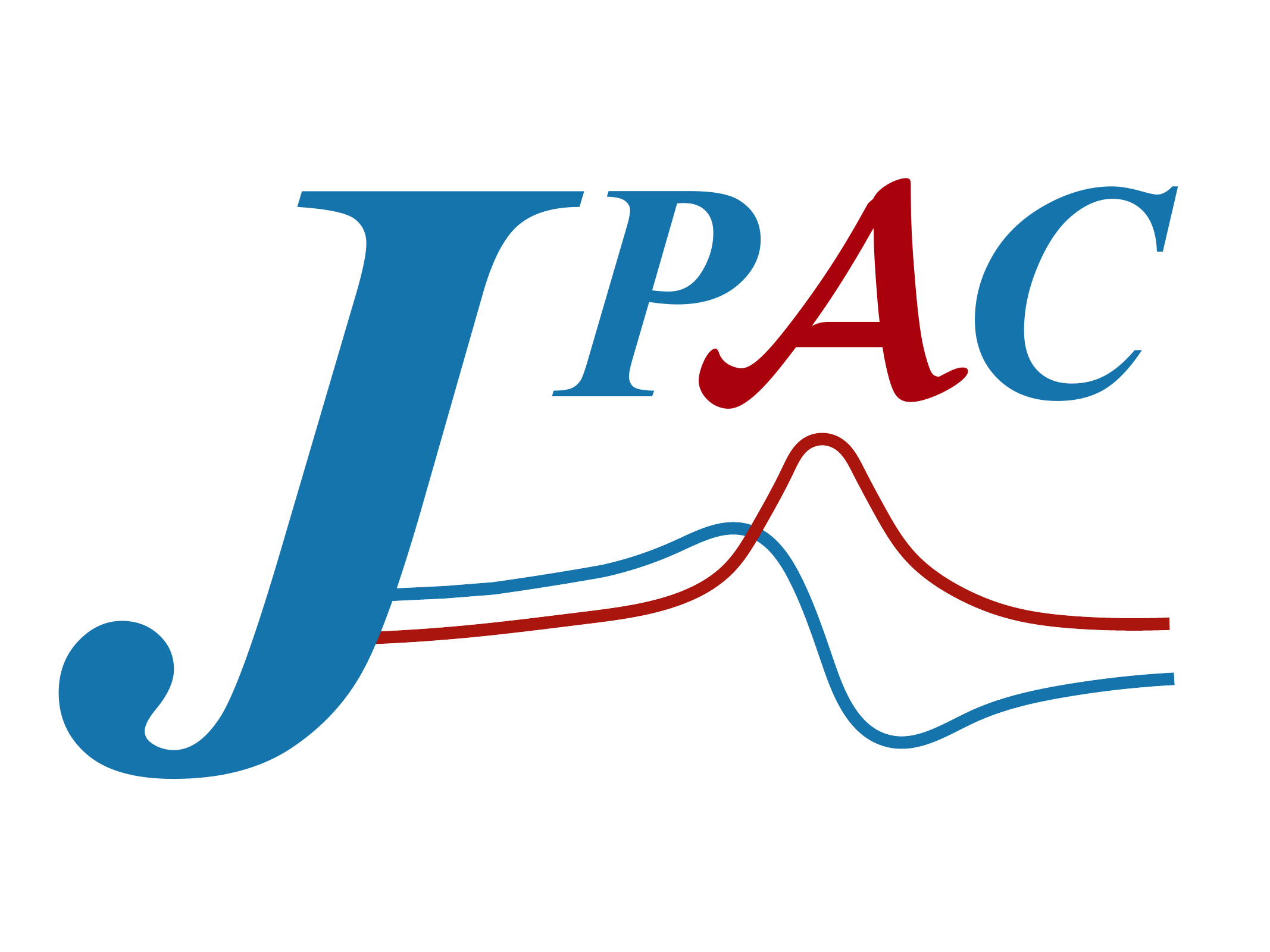}}
\preprint{JLAB-THY-25-4248}

\newcommand{\ific}{Instituto de F\'{i}sica Corpuscular, Centro Mixto Universidad de Valencia-CSIC, Institutos de Investigaci\'{o}n de Paterna, Aptdo. 22085, E-46071 Valencia, Spain}

\newcommand{\ub}{Departament de F\'isica Qu\`{a}ntica i Astrof\'isica (FQA), Universitat de Barcelona (UB), c.Mart\'i i Franqu\`{e}s, 1, 08028 Barcelona, Spain}
\newcommand{\iccub}{Institut de Ci\`{e}ncies del Cosmos (ICCUB), Universitat de Barcelona (UB), c.Mart\'i i Franqu\`{e}s, 1, 08028 Barcelona, Spain}

\newcommand{\ceem}{Center for  Exploration  of  Energy  and  Matter,
Indiana  University,
Bloomington,  IN  47403,  USA}
\newcommand{\indiana}{Department of Physics,
Indiana  University,
Bloomington,  IN  47405,  USA}
\newcommand{\jlab}{Theory Center,
Thomas  Jefferson  National  Accelerator  Facility,
Newport  News,  VA  23606,  USA}
\newcommand{\icn}{Instituto de Ciencias Nucleares, 
Universidad Nacional Aut\'onoma de M\'exico, Ciudad de M\'exico 04510, Mexico}
\newcommand{\hiskp}{Universit\"at Bonn,
Helmholtz-Institut f\"ur Strahlen- und Kernphysik, 53115 Bonn, Germany}
\newcommand{\catania}{INFN Sezione di Catania, I-95123 Catania, Italy}

\newcommand{\messina}{Dipartimento di Scienze Matematiche e Informatiche, Scienze Fisiche e Scienze della Terra, Universit\`a degli Studi di Messina, I-98166 Messina, Italy}
\newcommand{\cern}{CERN, 1211 Geneva 23, Switzerland}
\newcommand{\ucm}{Departamento de F\'isica Te\'orica, Universidad Complutense de Madrid and IPARCOS, 28040 Madrid, Spain}
\newcommand{\mainz}{Institut f\"ur Kernphysik \& PRISMA$^+$  Cluster of Excellence, Johannes Gutenberg Universit\"at,  D-55099 Mainz, Germany}

\newcommand{\uw}{Department of Physics, University of Washington, Seattle, WA 98195, USA}

\author[a]{A.Garc\'ia-Lorenzo\orcidlink{0009-0001-3200-9996}}
\emailAdd{Amador.Garcia@ific.uv.es}
\author[a]{M.~Albaladejo\orcidlink{0000-0001-7340-9235}}
\emailAdd{Miguel.Albaladejo@ific.uv.es}
\author[b,c]{S.~Gonz\`{a}lez-Sol\'is\orcidlink{0000-0003-1947-5420}}
\emailAdd{sergig@icc.ub.edu}
\author[b,c]{N.~Hammoud\orcidlink{0000-0002-8395-0647}}
\author[b,c]{V.~Mathieu\orcidlink{0000-0003-4955-3311}}
\author[d]{G.~Monta\~na\orcidlink{0000-0001-8093-6682}}
\author[e,f]{A.~Pilloni\orcidlink{0000-0003-4257-0928}}
\author[g]{D.~Winney\orcidlink{0000-0002-8076-243X}}
\author[d,h,i]{A.P.~Szczepaniak\orcidlink{0000-0002-4156-5492}}
\affiliation[a]{\ific}
\affiliation[b]{\ub}
\affiliation[c]{\iccub}
\affiliation[d]{\jlab}
\affiliation[e]{\messina}
\affiliation[f]{\catania}
\affiliation[g]{\hiskp}
\affiliation[h]{\indiana}
\affiliation[i]{\uw}

\abstract{This work studies the $\phi \to 3\pi$ decay and the $\phi \to \pi^0 \gamma^\ast$ transition form factor, utilizing the Khuri-Treiman formalism to account for analyticity, crossing, and unitarity. Using once-subtracted dispersion relations, we perform a simultaneous fit to the $\phi \to 3\pi$ Dalitz plot distribution and the $\phi \to \pi^0 \gamma^\ast$ measurements from the KLOE collaboration, finding good agreement with these experimental data. These results reaffirm the applicability of the Khuri-Treiman approach in the analysis of three-body decays. An interesting result is that the subtraction constant appearing in the equations is similar to a sum rule expectation, in contrast to analogous studies of $\omega \to 3\pi$ decays and $\omega \to \pi^0 \gamma^\ast$, which shows significant deviations. Our results also provide a reasonable description of the trend of the transition form factor data from BaBar in the $e^{+}e^{-}\to\phi\pi^{0}$ scattering region.
These intriguing theoretical differences between the decays of $\phi$ and $\omega$ could encourage further experimental measurements to assess the discrepancies and refine the theoretical predictions.}

\frenchspacing
\maketitle


\section{Introduction}\label{sec:introduction}

Accurately characterizing the amplitudes of three-particle final states remains an open challenge in hadron physics, which has gained increasing relevance with the advent of high-precision measurements from experiments such as GlueX, KLOE, and BaBar \cite{Ghoul:2015ifw,KLOE:2003kas,KLOE-2:2016pnx,BaBar:2007ceh,Pacetti:2009pg}. These experiments have identified, or aim to identify, various exotic states that decay into three particles. A precise understanding of three-body amplitudes is also crucial for determining resonance parameters in lattice QCD studies~\cite{Briceno:2017max,Jackura:2019bmu,Briceno:2019muc,Mai:2019pqr,Culver:2019vvu}.  

The Khuri-Treiman (KT) equations~\cite{Khuri:1960zz} provide a robust framework for studying three-body decays at low energies, as they incorporate analyticity, unitarity and crossing symmetry through dispersion relations. This formalism avoids the limitations associated with finite truncations when expanding the amplitude in partial waves, at the expense of losing analyticity in angular momentum. KT-based approaches have been extensively applied to isospin-breaking decays such as $\eta \to 3\pi$ \cite{Guo:2014vya,Guo:2015zqa,Guo:2016wsi,Colangelo:2016jmc,Colangelo:2018jxw,Albaladejo:2017hhj, Gasser:2018qtg}, as well as numerous other reactions~\cite{Niecknig:2012sj, Danilkin:2014cra, Niecknig:2015ija, Isken:2017dkw, Niecknig:2017ylb}. The method was later extended to account for all spin, isospin, parity, and charge conjugation quantum numbers of the decaying particle~\cite{Albaladejo:2019huw} (see also Ref.\,\cite{Mikhasenko:2019rjf}).  

Among its various applications, the decay of light vector isoscalar resonances $\omega / \phi \to 3\pi$ serves as an important benchmark for testing dispersion theory. Due to Bose symmetry, only odd angular momentum contributions are allowed in the $\pi\pi$ channels, with the dominant one being the $J = I = 1$ isobar, namely the $\rho$ meson, which plays a key role in the long-established Gell-Mann--Sharp--Wagner (GSW) mechanism~\cite{Gell-Mann:1962hpq,Meissner:1987ge}. This mechanism connects the $J = I = 1$ $\pi\pi$ partial wave amplitude to the results obtained from Roy equation analyses of $\pi\pi$ scattering~\cite{GarciaMartin:2011cn}. Existing studies on the decays $\omega \to 3\pi$ and $\phi \to3\pi$~\cite{Niecknig:2012sj,Danilkin:2014cra,Ananthanarayan:2014pta,Caprini:2015wja,Dax:2018rvs} employ dispersion relations to compute the Dalitz plot (DP) distribution. When unsubtracted dispersion relations are used, they provide shape predictions that are independent of adjustable parameters. More recently, KT equations have been successfully applied to the vector charmonium decay $J/\psi \to 3\pi$~\cite{JPAC:2023nhq} experimental data from BESIII~\cite{BESIII:2012vmy}. 

There have also been predictions of the $\phi \to \pi^0 \gamma^{\ast}$ transition form factor (TFF) \cite{Schneider:2012ez,Danilkin:2014cra}, using the $\phi \to 3\pi$ amplitude as input. However, the measurement of the $\phi \to \pi^0\gamma^{\ast}$ TFF by KLOE~\cite{KLOE-2:2016pnx} was only available after these studies were published. This TFF is closely related to the $\phi \to 3\pi$ amplitude at low energies. In this work, we explore this connection by introducing an additional subtraction term with a free complex parameter capable of absorbing effects beyond elastic unitarity, similarly as done in Ref.\,\cite{JPAC:2020umo}. We will perform a simultaneous fit to the $\phi \to 3\pi$ DP and the $\phi \to \pi^0\gamma^{*}$ TFF, which is the main novelty of this work respect to previous ones. In particular, we will demonstrate the relevance of a proper inclusion of crossed-channel effects through KT equations by comparing the DP shape with a simpler model. Also, as a result of our analysis, we reaffirm previous results concerning exclusively the $\phi \to 3\pi$ decay~\cite{Niecknig:2012sj}.

Another key motivation for this study is to compare the decays $\phi \to 3\pi$ and $\omega \to 3\pi$. These two decays are expected to be very similar. However, this expectation is challenged by the results of previous analyses~\cite{Niecknig:2012sj,JPAC:2020umo}. In both cases, subtracted dispersion relations were employed, but the subtraction constant required for the analysis of $\omega \to 3\pi$ in Ref.~\cite{JPAC:2020umo} significantly deviates from the expected sum-rule value, much more than in the case of $\phi \to 3\pi$~\cite{Niecknig:2012sj}. One possible explanation pointed out in Ref.~\cite{JPAC:2020umo} is the notably large values of the $\omega \to \pi^0\gamma^{\ast}$ TFF reported by the NA60 Collaboration~\cite{Arnaldi:2009aa,Arnaldi:2016pzu} for $\pi\pi$ invariant masses around $0.6\,\text{GeV}$. As a consequence of this, the description of the $\omega\to\pi^{0}\gamma^{*}$ at high energies was incompatible with the experimental data from the CMD and SND Collaborations~\cite{CMD-2:2003bgh,Achasov:2013btb} extracted from $e^{+}e^{-}\to\omega\pi^{0}$ scattering~\cite{JPAC:2020umo}.
In contrast, the data from the MAMI Collaboration~\cite{Adlarson:2016hpp} show no such enhancement, albeit with larger uncertainties, and was found to be compatible with the high energy data from the scattering region under the same description~\cite{JPAC:2020umo}. Thus, a renewed study of the $\phi \to 3\pi$ DP in conjunction with the TFF data is timely and necessary to further investigate this discrepancy. As we will see, the $\phi\to\pi^{0}\gamma^{*}$ transition form factor resulting from our analysis not only provides a satisfactory description of the low energy data from KLOE but also the trend of the high energy data from BaBar~\cite{BaBar:2007ceh} in the $e^{+}e^{-}\to\phi\pi^{0}$ scattering region, which is another outcome of this work.

The paper is organized as follows. In Section~\ref{sec:formalism}, we briefly review the KT formalism for $\phi \to 3\pi$ decays, discuss the kinematics, and explore its connection to the $\phi\pi^0$ TFF. Section~\ref{sec:results} presents the fits to KLOE data along with a discussion of our fitting approach. Finally, our conclusions are summarized in Section~\ref{sec:summary}.

\section{Formalism}\label{sec:formalism}

\subsection{Kinematics}\label{subsec:kin}

We begin by defining the terms and conventions that will be used throughout this text, as well as examining the kinematic implications of the decay process:
\begin{equation}
\phi(k_1) \to \pi^{+}(k_2)\pi^{0}(k_3)\pi^{-}(k_4)\,.
\label{Reaction}
\end{equation}
where $k_i^{\mu}=\left(E_i, \vec{k}_i \right)$ represents the four-momentum of the particles involved.

While our main focus centers on the physical decay region, it is simpler to express amplitudes within the framework of scattering, transforming one of the outgoing pions with $k_i^{\mu}$ into its ingoing antiparticle by changing $k^{\mu}_i \to -k^{\mu}_i$. In this manner, we now have three scattering processes:
\begin{equation}
    \begin{aligned}
        \text{$s$-channel:} \quad & \phi(k_1)\pi^{0}(-k_3) \to \pi^{+}(k_2)\pi^{-}(k_4)\,,\\
        \text{$t$-channel:} \quad & \phi(k_1)\pi^{-}(-k_2) \to \pi^{0}(k_3)\pi^{-}(k_4)\,,\\
        \text{$u$-channel:} \quad & \phi(k_1)\pi^{+}(-k_4) \to \pi^{+}(k_2)\pi^{0}(k_3)\,.\\
    \end{aligned}
\end{equation}
The three processes are identical under the limit of isospin symmetry. Therefore, we will now focus on the $s$-channel by adopting the center of mass reference frame of the two charged pions $\pi^{+}\pi^{-}$, aligning the $z$-axis with the direction of momentum of the incoming particles, which takes the form:
\begin{equation}
    k_1 = \left(E_{\phi}(s), +p(s)\;\hat{z}\right) \,,\;\;\; -k_3=\left(E_{\pi^{0}}(s), -p(s)\;\hat{z}\right)\,,
    \label{eq:inc-moment}
\end{equation}
with $p(s)$ the center-of-mass (CoM) momentum and $\hat{z}=\left(0,0,1\right)$. Similarly, the outgoing charged pions have:
\begin{equation}
    k_2 = \left(E_{\pi^{\pm}}(s), +q(s)\;\hat{q}_s\right) \,,\;\;\; k_4=\left(E_{\pi^{\pm}}(s), -q(s) \;\hat{q}_s\right)\,,
    \label{eq:out-moment}
\end{equation}
where $q(s)=|\bold{q}|$ is the modulus of the outgoing momentum and the unit vector $\hat{q}_s$ is defined as:
\begin{equation}
    \hat{q}_s = \left( \sin\theta_s, 0 , \cos \theta_s \right)\,.
\end{equation}
The Mandelstam variables are customarily defined as:
\begin{subequations}\label{eq:mandelstam}
    \begin{align}
        s & = (k_1+k_3)^2 = (k_2+k_4)^2\,, \\
        t & = (k_1-k_2)^2 = (k_4-k_3)^2\,, \\
        u & = (k_1-k_4)^2 = (k_3-k_2)^2\,, \\
    s + t + u & = \sum_{i}k_{i}^{\mu}k_{{i\mu}}=M^2_{\phi}+m^2_{\pi_0}+2m^2_{\pi^{\pm}} \equiv \Sigma\,. \label{eq:sum_mandelstam}
    \end{align}
\end{subequations} 
The three-momenta $p(s)$ and $q(s)$ are then given by:
\begin{subequations}\label{eq:modulus}
\begin{align}
    p(s) & = \frac{\lambda^{\frac{1}{2}}(s,M^2_{\phi},m^2_{\pi^0})}{2\sqrt{s}}\,,\\
    q(s) & = \frac{\lambda^{\frac{1}{2}}(s,m^2_{\pi^{\pm}},m^2_{\pi^\pm})}{2\sqrt{s}}\,,
\end{align}
\end{subequations}
where we have introduced the K\"all\'en (or triangle) function defined by \cite{Kallen:1964lxa}:
\begin{equation}
    \lambda(x,y,z)=x^2+y^2+z^2-2(xy+xz+yx)\,.
    \label{eq:kallen}
\end{equation}
For completeness and later reference, we show here the meson energies in the $\phi\pi^0$ CoM system, which are:
\begin{subequations}\label{eq:phipi-energy}
\begin{align}
    E_{\phi}(s)  & = \frac{s+M^2_{\phi}-m_{\pi^0}^2}{2\sqrt{s}}\,,\\
    E_{\pi^0}(s) & = \frac{s+m^2_{\pi^0}-M_{\phi}^2}{2\sqrt{s}}\,,\\
    E_{\pi^{\pm}}(s) & = \frac{\sqrt{s}}{2}\,.
    \end{align}
\end{subequations}
The cosine of the scattering angle $\theta_s$ is obtained by means of the dot product between Eq.\,\eqref{eq:inc-moment} and Eq.\,~\eqref{eq:out-moment},  giving:
\begin{equation}
    \cos \theta_s(s,t)= \frac{2t+s-\Sigma}{4p(s)q(s)}\,.
    \label{eq:cos}
\end{equation}
To obtain $\sin\theta_s$, let us consider the axial four-vector known as Kibble vector \cite{Kibble:1960zz}:
\begin{equation}
    K_\mu = \epsilon_{\mu\nu\rho\sigma}k_2^{\nu}k_{3}^{\rho}k_4^{\sigma}\,,
    \label{eq:Kibble-vector}
\end{equation}
where $\epsilon$ is the Levi-Civita symbol which is totally antisymmetric in all indices. Noting that one can take $K_\mu = \epsilon_{\mu\nu\rho\sigma}(k_2^{\nu}+k_4^{\nu}) k_3^{\rho} k_4^{\sigma}$, it follows that in the $s$-channel CoM frame:\footnote{Note that there is freedom in choosing the direction of the $y$-axis, which is chosen such that the momentum of the $\pi^+(k_2)$ has a positive $y$ component.}
\begin{equation}
    K_0 = 0\,, \qquad \vec{K} = \sqrt{s}p(s)q(s) \sin \theta_s\,\hat{y}\,.
    \label{eq:K-components}
\end{equation}
Therefore, we have:
\begin{equation}
    \phi(s,t) \equiv -4K_\mu K^\mu=4sp^2(s)q^2(s)\sin^2\theta_s\,,
    \label{eq:Kibble-def}
\end{equation}
where we have introduced the invariant function $\phi(s,t)$ known as Kibble function. Its value can be computed through the Gram determinant:
\begin{equation*}
    \phi(s,t) = -4K_\mu K^\mu= 4 
    \begin{vmatrix}
    k_{3\mu}^2  & k_{3\mu}k_2^\mu & k_{3\mu}k_4^{\mu} \\
    k_{2\mu}k_3^\mu & k_{2\mu}^2 & k_{2\mu}k_4^{\mu} \\
    k_{4\mu}k_3^\mu & k_{4\mu}k_2^{\mu} & k_{4\mu}^2 \\
    \end{vmatrix}\,,
\end{equation*}
which gives:
\begin{equation}
    \phi(s,t)=stu -s\left(M^2_{\phi}-m^2_{\pi^\pm}\right)\left(m^2_{\pi^0}-m^2_{\pi^\pm}\right)-m^2_{\pi^\pm}\left(M^2_{\phi}-m^2_{\pi^0}\right)^2\,.
    \label{eq:Kibble}
\end{equation}
Note that the boundaries of the physical regions of the process are defined through $\phi(s,t)=0$.
Finally, using Eq.\,\eqref{eq:Kibble-def} we obtain $\sin\theta_s$ as:
\begin{equation}
    \sin \theta_s = \frac{\sqrt{\phi(s,t)}}{2\sqrt{s}p(s)q(s)}\,.
    \label{eq:sin}
\end{equation}
There is a small symmetry breaking between the $s$- and $t,u$- channels due to the mass difference $\Delta m = m_{\pi^\pm}-m_{\pi^0} \approx (139.57 - 134.98)\,\text{MeV} = 4.59\,\text{MeV}$, which is practically irrelevant for the development of amplitudes using Khuri-Treiman equations.\footnote{While isospin breaking due to pion masses is not relevant, we will take into account in an effective manner the isospin breaking due to the appearance of $\rho^0$-$\omega$ mixing in the $\pi^+\pi^-$ channel, which is essential to properly describe the $\phi \to \pi^+\pi^-\pi^0$ DP.} Hence, for the rest of the manuscript, we will work in the isospin limit $m_{\pi^0}=m_{\pi^{\pm}}$, where we take as reference the mass of the charged pion $m\equiv m_{\pi^\pm}$, which allows us to take the results of the $s$-channel as a representative of the three channels.

\subsection[$\phi \to 3\pi$ amplitude from Khuri--Treiman equations]{\boldmath $\phi \to 3\pi$ amplitude from Khuri--Treiman equations}\label{subsec:KT}

The amplitude of our process Eq.\,\eqref{Reaction} can be given in terms of helicity amplitudes $\mathcal{H}_{\lambda}(s,t,u)$, which, in turn, depend on a Lorentz-invariant amplitude $F(s,t,u)$:
\begin{equation}
    \mathcal{H}_{\lambda}(s,t,u) = i \epsilon_{\lambda}^{\mu}(k_1) K_\mu F(s,t,u)\,,
    \label{eq:helam_1}
\end{equation}
where $\epsilon_{\lambda}^{\mu}$ is the polarization tensor of the $\phi$ meson with helicity $\lambda$ and $K_{\mu}$ is the Kibble vector in Eq.\,\eqref{eq:Kibble-vector}.
The $\phi$ meson has three possible helicity states $\lambda=-1,0,+1$. Thus, the differential decay width $\frac{d^2\Gamma}{dsdt}(s,t)$ is calculated by averaging over the three helicity states:
\begin{equation}
     \frac{d^2\Gamma}{dsdt}(s,t) = \frac{1}{(2\pi)^3}\frac{1}{32M^3_{\phi}}|M(s,t)|^2 \,,
     \label{eq:diff-decay}
\end{equation}
where:
\begin{equation}
    |M(s,t)|^2 = \frac{1}{3}\sum_{\lambda}|\mathcal{H}_{\lambda}|^2=\frac{1}{3}\sum_{\lambda}\epsilon^{\mu}_{\lambda}(k_1)K_{\mu}\ \left[K_{\nu} \epsilon^{\nu}_{\lambda}(k_1)\right]^{*} |F(s,t,u)|^2 \,.
    \label{eq:M}
\end{equation}
The right-hand side of Eq.\,\eqref{eq:M} can be computed using:
\begin{equation}
\sum_{\lambda}\epsilon^{\mu}_{\lambda}(k_1)\left(\epsilon^{\nu}_{\lambda}(k_1)\right)^* = -g^{\mu\nu}+\frac{k_1^{\mu}k_1^\nu}{M^2_{\phi}}\,,
\end{equation}
so that Eq.\,\eqref{eq:M} becomes:
\begin{equation}
    \left\lvert M(s,t) \right\rvert^2 = \frac{1}{3}(-K_{\mu}K^{\mu})|F(s,t,u)|^2=\frac{1}{3}\frac{\phi(s,t)}{4}|F(s,t,u)|^2\,,
\end{equation}
where we have used the fact that $k_1^{\mu}K_{\mu}=0$. 
Now, the Kibble function Eq.\,\eqref{eq:Kibble} must be computed in the isospin limit:
\begin{equation}
    \phi(s,t)=stu-m^2(M_{\phi}^2-m^2)^2\,,
\end{equation}
which gives us the limits for $s$ in the physical decay region by solving $\phi(s,t,u)=0$:
\begin{equation}
    s_{\text{min}}=4m^2 \,,\qquad s_{\text{max}}=(M_{\phi}-m)^2\,,
\end{equation}
while the limits of $t(s)$  can be obtained from Eq.\,\eqref{eq:cos} taking $\cos\theta_s=\pm 1$:
\begin{equation}
    t_{\pm}(s)=\frac{1}{2}\left[M^2_{\phi}+3m^2-s \right] \pm 2p(s)q(s)\,.
    \label{eq_tlim}
\end{equation}
Note that all the dynamics of the decay problem are encoded in the function $F(s,t,u)$. Its form can be deduced considering that the helicity amplitude Eq.\,\eqref{eq:helam_1} can be decomposed into helicity partial waves amplitudes as:
\begin{equation}
    \mathcal{H}_{\lambda}(s,t,u) = \sum_{J \; \text{odd}}^{\infty}(2J+1)d^J_{\lambda 0}(\theta_s)h^J_{\lambda}(s)\,,
\end{equation}
where $d^J_{\lambda 0}$ are the Wigner $d$-functions and $\theta_s$ is the angle given by Eqs.\,\eqref{eq:cos} and \eqref{eq:sin}. For our decay $\phi \to 3\pi$, $\mathcal{H}_0=0$ and $\mathcal{H}_{+}=\mathcal{H}_{-}$, due to parity. This formula allows to write the partial wave expansion for the invariant amplitude $F(s,t,u)$ as:
\begin{equation}
    F(s,t,u)=\sum_{J \;\text{odd}}^{\infty}\left(p(s)q(s) \right)^{J-1}P'_{J}(\cos\theta_s)f_J(s)\,,
    \label{eq:pwFstu}
\end{equation}
where $P'_{J}(z)$ is the derivative of the Legendre polynomial, and with $f_J(s)$ the kinematic-singularity-free partial wave amplitude for total angular momentum $J$ of the two-pion system, which has the form:
\begin{equation}
    f_J(s) \equiv \sqrt{\frac{2}{s}}\frac{2J+1}{\sqrt{J(J+1)}}\frac{h_+^{J}(s)}{\left(p(s)q(s)\right)^J}\,.
\end{equation}
In general, Eq.\,(\ref{eq:pwFstu}) is an infinite sum of partial waves with the expansion performed in the $s$-channel, and with $f_J(s)$ carrying both left- and right-hand cuts (LHC and RHC, respectively). In practice, however, a truncation to a finite number of partial waves is necessary when modeling processes at low energies. Such truncation introduces a violation of the analytical properties of the crossed $t$ and $u$ channels. The KT formalism partially recovers these properties by replacing the infinite sum of partial waves in the $s$-channel with three truncated sums of single-Mandelstam-variable functions, the so-called isobar amplitudes, one for each $s,t$ and $u$ channels. The KT representation of the amplitude then reads:
\begin{equation}
\begin{aligned}
    F(s,t,u) = &  
    \sum_{J\; \text{odd}}^{J_\text{max}} (2J+1)P'_J(\cos\theta_s)F_J(s) + \\
  & \sum_{J\; \text{odd}}^{J_\text{max}} (2J+1)P'_J(\cos\theta_t)F_J(t) + 
    \sum_{J\; \text{odd}}^{J_\text{max}} (2J+1)P'_J(\cos\theta_u)F_J(u)\,.
\end{aligned}
\end{equation}
Note that because the amplitude $F(s,t,u)$ must be symmetric in $s$, $t$, and $u$, the truncation is performed at the same $J_\text{max}$ for each isobar expansion. Truncating each sum at $J_{\text{max}}=1$ (and omitting the subindex $J$ in $F_J(s)$), we obtain the following isobar decomposition~\cite{Niecknig:2012sj,Danilkin:2014cra,Albaladejo:2019huw,JPAC:2023nhq}:
\begin{equation}
    F(s,t,u)=F(s)+F(t)+F(u)\,,
    \label{eq:Fstu}
\end{equation}
where each isobar amplitude, $F(x)$, has only a RHC in its corresponding Mandelstam variable. In the previous section, we demonstrated that the three channels are equivalent in the isospin limit, so we only need to calculate $F(s)$. Projecting Eq.\,\eqref{eq:Fstu} into the $s$-channel partial wave, we obtain the relation between $f_1(s)$ and $F(s)$:
\begin{equation}
    f_1(s)=F(s)+\widehat{F}(s)\,,
    \label{eq:f1}
\end{equation}
where:
\begin{equation}
    \widehat{F}(s) \equiv \frac{3}{2}\int_{-1}^1 (1-z^2_s)F(t(s,z_s)) \;dz_s\,,
    \label{eq:F1}
\end{equation}
is known as the inhomogeneity and contains the $s$-channel projection of the contributions due to the $t$- and $u$-channels. 

The contribution to the generalized unitarity relation for $F(s)$ from $\pi\pi$ intermediate states reads \cite{Albaladejo:2017hhj,Mandelstam:1960zz}: 
\begin{equation}
    \Delta F(s) \equiv F(s+i\epsilon)-F(s-i\epsilon) 
    =2i(F(s)+\widehat{F}(s))\sin\delta(s)e^{-i\delta(s)}\theta(s-4m^2)\,,
    \label{eq:Disc}
\end{equation}
where $\delta(s)$ is the $P$-wave $\pi\pi$ phase shift. Note that we have assumed that there are only two-pion intermediate states for the $\phi\to 3\pi$ decay. An unsubtracted dispersion relation (UDR) for $F(s)$ can be written from Eq.\,\eqref{eq:Disc} as:
\begin{equation}
    F(s)=\frac{1}{2\pi i}\int_{4m^2}^{\infty}\frac{\Delta F(s')}{s'-s}\,d s'\,.
\end{equation}
Its solution can be written as:
\begin{equation}
F(s)=\Omega(s)\left(a+\frac{s}{\pi}\int_{4m^2}^{\infty}\frac{ds'}{s'}\frac{\sin\delta(s')\widehat{F}(s')}{\left\lvert \Omega(s') \right\rvert (s'-s)} \right)\,,
    \label{eq:USR}
\end{equation}
where $\Omega(s)$ is the well-known Omnès function \cite{Omnes:1958hv}:
\begin{equation}
    \Omega(s)=\exp\left[
    \frac{s}{\pi}\int_{4m^2}^{\infty}\frac{\delta(s')}{s'(s'-s)}ds' 
    \right]\,,
\end{equation}
which is also the solution of Eq.\,\eqref{eq:Disc} taking $\widehat{F}(s)=0$, hence the name inhomogeneity for $\widehat{F}(s)$. The subtraction constant $a$ is in general complex, {\it{i.e}} $a=|a|e^{i\phi_{a}}$. While $|a|$ can be fixed from the measured $\phi\to3\pi$ decay width, no observable is sensitive to $\phi_{a}$. Therefore, we can simply set $\phi_{a}=0$ and take $a$ to be a real overall normalization of the amplitude.

\begin{figure}[t]\centering
    \includegraphics[height=11cm,keepaspectratio]{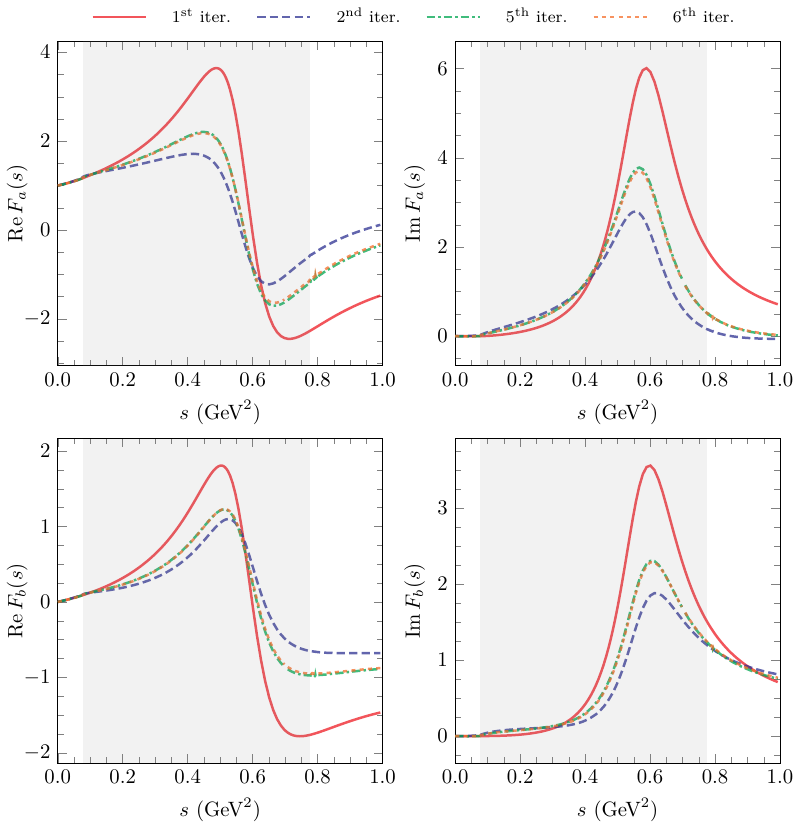}
    \caption{Iterative solutions of the KT equations for the $F_a(s)$ and $F_b(s)$ functions [\textit{cf.} Eqs.\,\eqref{eq:FaFb_def}]. The shaded area represents the physical $\phi \to 3\pi$ decay region, \textit{i.e.}, $4m_{\pi}^2 \leqslant s \leqslant (m_{\phi}-m_{\pi})^2$. Note that the $\rho$ peak lies clearly within this range.\label{fig:iterations}}
\end{figure}

One can also write a once-subtracted dispersion relation (ODR) for $F(s)$ from Eq.\,\eqref{eq:Disc} as:
\begin{equation}
    F(s)=F(0)+\frac{s}{2\pi i}\int_{4m^2}^{\infty}\frac{\Delta F(s')}{s'(s'-s)}ds'
    \label{ODR}\,,
\end{equation}
with solution:
\begin{equation}
F(s)=\Omega(s)\left(a+b' s+\frac{s^2}{\pi}\int_{4m^2}^{\infty}\frac{ds'}{s'^2}\frac{\sin\delta(s')\widehat{F}(s')}{\left\lvert \Omega(s') \right\rvert (s'-s)} \right)
\label{eq:OSR}\,,
\end{equation}
The subtraction constants $a$ and $b'$ enter linearly into Eq.~\eqref{eq:OSR}, and hence $F(s)$ can be decomposed into a linear combination of two functions $F_a(s)$ and $F_b(s)$:
\begin{equation}
    F(s)=a\,F_a(s) + b'\,F_b(s) = a \left( F_a(s) + b\,F_b(s) \right)\,,
    \label{eq:linearcomb}
\end{equation}
where $b=b'/a$. The $F_a$ and $F_b$ functions are:
\begin{subequations}\label{eq:FaFb_def}
\begin{align}
F_a(s)=\Omega(s)\left(1+\frac{s^2}{\pi}\int_{4m^2}^{\infty}\frac{1}{(s')^2}\frac{\sin\delta(s')\hat{F_a}(s')}{|\Omega(s')|(s'-s)}ds'\right)\,, \\  F_b(s)=\Omega(s)\left(s+\frac{s^2}{\pi}\int_{4m^2}^{\infty}\frac{1}{(s')^2}\frac{\sin\delta(s')\hat{F_b}(s')}{|\Omega(s')|(s'-s)}ds'\right)\,,
\end{align}
\end{subequations}
which are independent solutions of Eq.\,\eqref{ODR}. In Eqs.\,\eqref{eq:FaFb_def}, the functions $\widehat{F}_{a,b}(s)$ are obtained as in Eq.\,\eqref{eq:F1} with the change $F(s) \to F_{a,b}(s)$. The advantage of using these functions is that they only need to be calculated once since they do not depend on $a$ and $b$. They can be solved iteratively, with rapid convergence, as shown in Fig.\,\ref{fig:iterations}.\footnote{The phase shifts $\delta(s)$ used as inputs are discussed below in Sec.\,\ref{subsec:globalfits}.} Additionally, introducing a subtraction reduces sensitivity to the high energy region, where the phase shift is unknown. It is also important to note that, by fitting $b$ to experimental data, we can reproduce behaviors that have not been explicitly addressed when developing the KT equations. Finally, we observe that the ODR can be reduced to the UDR by setting $b$ to the following value:
\begin{equation}
    b_{\text{sum}} \equiv \frac{1}{\pi}\int_{4m^2}^{\infty}\frac{1}{(s')^2}\frac{\sin\delta(s')\hat{F}(s')}{|\Omega(s')|}ds' = \frac{x_a}{1-x_b}\,,
    \label{eq:sumrule}
\end{equation}
with:
\begin{equation}
x_{\{a,b\}} = \frac{1}{\pi} \int ds' \frac{\hat{F}_{\{a,b\}}(s')}{{s'}^2} \frac{\sin\delta(s')}{\left\lvert \Omega(s') \right\rvert}\,.
\end{equation}

Additionally, due to the high mass of the $\phi$ meson and to the isospin breaking effects, in the $s$-channel there is mixing between the $\rho^0$ and the $\omega$ mesons. Therefore, we have to introduce an additional term in the $F(s)$ isobar of Eq.\,\eqref{eq:Fstu}, to account for the $\omega$, in the following way~\cite{KLOE:2003kas,Niecknig:2012sj}:
\begin{equation}
    F(s) \;\; \to \;\; F(s) +A\frac{M^2_\omega}{M^2_{\omega}-s-i\sqrt{s}\Gamma_{\omega}}\,,
    \label{eq:resomega}
\end{equation}
where $A$ is a complex coupling constant. This term is added after the iterative solution is obtained.

\subsection[$\phi \to \pi^0\gamma^*$ transition form factor]{\boldmath $\phi \to\pi^0\gamma^*$ transition form factor}\label{subsec:FF}

We introduce here the $\phi\pi^0$ TFF $f_{\phi\pi^0}(s)$, which controls the $\phi \to \pi^0\gamma^*$ amplitude. The dispersive representations of $f_{\phi\pi^0}(s)$ are determined by the following unitarity relation:
\begin{equation}
    \Delta f_{\phi\pi^0}(s) \equiv f_{\phi\pi^0}(s+i\epsilon)-f_{\phi\pi^0}(s-i\epsilon)=i\frac{p^3(s)}{6\pi\sqrt{s}} {F^{V}_{\pi}(s)}^{\ast} f_1(s)\theta(s-4m^2)\,,
    \label{eq:discFF}
\end{equation}
where one can see that the discontinuity only affects the RHC. Only the two charged pions annihilation has been taken into account in Eq.\,\eqref{eq:discFF}, where $f_1(s)$ can be obtained from KT equations as in Eqs.\,\eqref{eq:f1} and \eqref{eq:F1}. Here, in addition, one can approximate the pion vector form factor $F^V_{\pi}(s)$ by the Omnès function $\Omega(s)$ because of the relatively low invariant mass of the $\phi\pi^0$ system explored in this work. In analogy to and for the sake of comparison with Ref.\,\cite{JPAC:2020umo}, we write down a once-subtracted dispersion relation for $f_{\phi\pi^0}(s)$ as:
\begin{equation}
    f_{\phi\pi^0}(s)=\left\lvert f_{\phi\pi^0}(0) \right\rvert e^{i\theta_{\phi\pi^0}(0)}+\frac{s}{12\pi^2}\int_{4m^2}^{\infty}\frac{1}{\sqrt{(s')^3}}\frac{p^3(s'){\Omega(s')}^{\ast} f_{1}(s')}{(s'-s)}ds'\,,
    \label{eq:Transition}
\end{equation}
where the complex phase $\theta_{\phi\pi^0}(0)$ generally cannot vanish because we have imposed that the overall normalization constant $a$ in the amplitude in Eq.\,\eqref{eq:USR} is real. Finally, we note that the modulus of the TFF at zero $|f_{\phi\pi^0}(0)|$ is directly related to the experimental decay width for $\phi \to \pi^0\gamma$: 
\begin{equation}
    \Gamma_{\phi \to \pi^0\gamma} = \frac{e^2}{4\pi}\frac{\left(M^2_\phi-m^2_{\pi^0}\right)^3}{24M^3_{\phi}}\left\lvert f_{\phi\pi^0}(0) \right\rvert^2\,.
    \label{eq:decayFF}
\end{equation}

\section{Results}\label{sec:results}

\subsection{Experimental information}\label{subsec:general}
In our approach we have a total of eight real parameters to fit. From the $\phi \to 3\pi$ amplitude obtained from KT equations [\textit{cf.} Eq.\,\eqref{eq:linearcomb}] we have a real constant $a$ and a complex value $b=b_r +i b_i$. The additional $\omega$-$\rho^0$ mixing introduces another complex constant $A$ [\textit{cf.} Eq.\,\eqref{eq:resomega}]. We require another real parameter $N$, which serves as an adjustable parameter for the total number of events in the $\phi\to3\pi$ DP distribution (\textit{cf.} Eq.\,\eqref{eq:Nevents} below). Finally, from the TFF [\textit{cf.} Eq.\,\eqref{eq:Transition}], we obtain two additional parameters: $|f_{\phi\pi^{0} }(0)|$ and $\theta_{\phi\pi^0}(0)$. To determine these unknown constants, we will use the following experimental information:
\begin{enumerate}[label={\alph*)}]
\item Measurement of $\phi\to3\pi$ decay DP~\cite{KLOE:2003kas} and transition form factor $|f_{\phi\pi^0}(s)/f_{\phi\pi^0}(0)|^2$ by the KLOE collaboration~\cite{KLOE-2:2016pnx,Pacetti:2009pg}.
\item The decay widths $\phi \to 3\pi$ and $\phi \to \pi^0 \gamma^\ast$, for which we take the PDG values \cite{ParticleDataGroup:2024cfk}: $\Gamma_\phi = 4.249 \pm 0.013\,\MeV$, the branching ratios $\mathcal{B}(\phi \to 3\pi) = 15.4 \pm 0.4\ \%$ and $\mathcal{B}(\phi \to \pi^0 \gamma) = 0.132 \pm 0.05\ \%$.
\item The rest mass $M_{\omega}=782.66\,\MeV$ and the decay width $\Gamma_{\omega}=8.68\,\MeV$ of the PDG.
\end{enumerate}
To fit our parameters, we define a global $\chi^2$ function, which takes the form:
\begin{equation}
    \chi^2 \equiv \chi^2_{\text{DP}} + \chi^2_{\Gamma_{3\pi}}+\chi^2_{\text{TFF}}+\chi^2_{\Gamma_{\pi^0\gamma^\ast}}\,.
    \label{eq:globalfit}
\end{equation}
The function $\chi^2_{\text{DP}}$ refers to the fit of the DP distribution mentioned above, its expression being given by:
\begin{equation}   \chi^2_{\text{DP}}=\sum_{i}\left(\frac{N^{\text{exp}}_{\text{ev},i}-N^{\text{th}}_{\text{ev}, i}}{\sigma_{N^{\text{exp}}_{\text{ev},i}}} \right)^2\,.
\end{equation}
The DP is divided into bins, each one having the experimental number of events $N^{\text{exp}}_{\text{ev},i}$ (and its associated error $\sigma_{N^{\text{exp}}_{\text{ev},i}}$), measured in a square region of the phase space with semi-width $\Delta X$= $\Delta Y=8.75\,\text{MeV}$ and whose center is given by the parameters $X$ and $Y$ defined in terms of $s$ and $t$ by:
\begin{subequations}\label{eq:XYdefinition}
\begin{align}
    X & = \frac{2t+s-\Sigma}{2M_{\phi}}\,, \label{eq:Xdefinition}\\
    Y & =-\frac{s-(M_{\phi}-m_{\pi^0})^2}{2M_{\phi}}\,. \label{eq:Ydefinition}
\end{align}
\end{subequations}
In order to compute the number of events in each bin from the theoretical differential decay width Eq.\,\eqref{eq:diff-decay}, we write: 
\begin{equation}
    N^{\text{th}}_{\text{ev},i} = 
    N \left( \frac{\Gamma^{\text{th}}_{3\pi , i}}{\Gamma^{\text{th}}_{3\pi}} \right)
    =
    \frac{N}{(2\pi)^3 384 M^3_{\phi} \Gamma^{\text{th}}_{3\pi}}\int_{s_i-\Delta s}^{s_i+\Delta s}\int_{t_i-\frac{(s-s_i)-\Delta t}{2}}^{t_i+\frac{(s-s_i)+\Delta t}{2}}\phi(s,t)\left|F(s,t) \right|^2 \, dt\, ds \,,
    \label{eq:Nevents}
\end{equation}
where we have defined $\Delta s=\Delta t=8.75\, M_{\phi} = \Delta X\, M_\phi$. Note that each square bin in the $(X,Y)$-plane transforms into a parallelogram in the $(s,t)$-plane, as reflected in the $s$ and $t$ integration limits in Eq.\,\eqref{eq:Nevents}. Also note that, as stated above, we have introduced $N$, which corresponds to the total number of theoretical events, as a fitting parameter, not as an input. 

We also fit the $\phi \to 3\pi$ total decay width, which corresponds to the following integral over the whole phase space:
\begin{equation}
    \Gamma^\text{th}_{3\pi} = \frac{1}{(2\pi)^{3}384M^3_{\phi}}\int_{4m^2}^{(M_\phi-m)^2}\int_{t_{-}(s)}^{t_{+}(s)}\phi(s,t)\left|F(s,t) \right|^2 \, dt \, ds\,,
\end{equation}
for which fit we define the following $\chi^2_{\Gamma_{3\pi}}$ function:
\begin{equation}
    \chi^2_{\Gamma_{3\pi}} = \left( \frac{\Gamma^\text{exp}_{3\pi}-\Gamma^\text{th}_{3\pi}}{\sigma_{\Gamma^\text{exp}_{3\pi}}}\right)^2\,,
\end{equation}
with $\Gamma^\text{exp}_{3\pi}=\Gamma_{\phi} \, \mathcal{B}(\phi\to 3\pi)$ and $\sigma_{\Gamma^\text{exp}_{3\pi}}=\Gamma^\text{exp}_{3\pi} \, \sqrt{\left(\frac{\Delta \mathcal{B}(\phi\to 3\pi)}{\mathcal{B}(\phi\to 3\pi)}\right)^2+\left(\frac{\Delta\Gamma_{\phi}}{\Gamma_{\phi}}\right)^2}$.

To fit the TFF [\textit{cf.} Eq.\,\eqref{eq:Transition}], we define the $\chi^2$-square function $\chi^2_\text{TFF}$ as:
\begin{equation}
    \chi^2_{\text{TFF}} = \sum_{i} \left( \frac{\left|F_{\phi\pi^0}(s_i)\right|^{2}_\text{exp}-\left|F_{\phi\pi^0}(s_i)\right|^{2}_\text{th}}{\sigma^\text{TFF}_i}\right)^2 \,,
\end{equation}
where:
\begin{equation}
F_{\phi\pi^0}(s)=\frac{f_{\phi\pi^0}(s)}{f_{\phi\pi^0}(0)}
\end{equation}
is the normalized TFF. $\left|F_{\phi\pi^0}(s)\right|_\text{exp}$ and $\sigma^\text{TFF}_i$ are the experimental normalized TFF and its error, respectively, provided by the data in Ref.\,\cite{KLOE-2:2016pnx}. $\left|F_{\phi\pi^0}(s)\right|^2_\text{th}$ is the average in each range $\left[\sqrt{s}-\Delta \sqrt{s},\sqrt{s}+\Delta \sqrt{s}\right]$ of our normalized transition form factor:
\begin{equation}\label{eq:TFF_average}
    \left|F_{\phi\pi^0}(s_i)\right|^{2}_\text{th}=\frac{1}{4\sqrt{s_i}\Delta \sqrt{s_i}}\int_{(\sqrt{s_i}-\Delta \sqrt{s_i})^2}^{(\sqrt{s_i}+\Delta \sqrt{s_i})^2}\left|F_{\phi\pi^0}(s)\right|^{2} ds \,,
\end{equation}

Finally, we also fit the $\phi \to \pi^0 \gamma^\ast$ decay width, involving the quantity $f_{\phi\pi^0}(0)$, for which we define $\chi^2_{\Gamma_{\phi\pi^0}}$ as:
\begin{equation}
\chi^2_{\Gamma_{\phi\pi^0}}=\left(\frac{\Gamma^\text{exp}_{\phi\pi^0}-\Gamma^\text{th}_{\phi\pi^0}}{\sigma_{\Gamma^\text{exp}_{\phi\pi^0}}}\right)^2\,,
\end{equation}
with $\Gamma^\text{exp}_{\phi\pi^0}=\Gamma_{\phi}\mathcal{B}(\phi\to \pi^0\gamma^\ast)$, $\sigma_{\Gamma^\text{exp}_{\phi\pi^0}}=\Gamma^\text{exp}_{\phi\pi^0}\sqrt{\left(\frac{\Delta \mathcal{B}(\phi\to \pi^0\gamma^\ast)}{\mathcal{B}(\phi\to \pi^0\gamma^\ast)}\right)^2+\left(\frac{\Delta\Gamma_{\phi}}{\Gamma_{\phi}}\right)^2}$ and $\Gamma^\text{th}_{\phi\pi^0}$ given by Eq.\,\eqref{eq:decayFF}.

\subsection{Global fit results}\label{subsec:globalfits}

To perform minimization of the $\chi^2$ function [\textit{cf.} Eq.\,\eqref{eq:globalfit}], we make use of the library \texttt{Minuit2} of C++, with \texttt{Migrad}, \texttt{Minos} and \texttt{Covariance} as subroutines. The total number of points to fit is:
\begin{equation}
    N = N_{\text{DP}}+N_{\Gamma_{3\pi}}+N_{\text{TFF}}+N_{\Gamma_{\pi^0\gamma^\ast}}=1860+1+15+1=1877 \,,
\end{equation}
where we assume that all the associated $\chi^2$ functions carry equal weight in the overall fit. It is important to note that in the analysis of $\phi \to 3\pi$ decays, we have excluded bins close to the DP boundary to mitigate potential isospin-breaking effects. We achieved this by selecting from the KLOE data \cite{KLOE:2003kas} only those bins that meet the condition $\phi(s,t)>0$ at the four corners of each bin $(X_i,Y_i)$ and that satisfy $N_{\text{ev},i}^{\text{exp}}>0$. This results in a total number of bins reduced to $N_\text{DP} = 1860$.

As an input for the KT equations, we need the $P$-wave phase shift $\delta(s)$. We will use four different parameterizations, which are solutions of the Roy equations of Refs.\,\cite{Garcia-Martin:2011iqs} and \cite{Pelaez:2019eqa}, and are approximately valid up to $1.3$ and $2$ $\GeV$, respectively. More specifically, we will denote the parameterization from Ref.\,\cite{Garcia-Martin:2011iqs} as $\delta_1(s)$, while $\delta_{2,3,4}(s)$ will represent solutions I, II, and III from Ref.\,\cite{Pelaez:2019eqa}, respectively. This will help estimate the eventual systematic uncertainties in our approach. From the upper validity limit to $s\to\infty$, we will smoothly guide the function to $\pi$, as in \cite{JPAC:2023nhq} and \cite{Gonzalez-Solis:2019iod}, to achieve the expected asymptotic $1/s$ fall-off behavior for the pion vector form factor, which is approximated here by the Omnès function $\Omega(s)$ as discussed above.

\begin{table}\centering
\begin{tabular}{|r|c|c|c|c|} \hline
 &                                               $\delta_1$  & $\delta_2$ & $\delta_3$ & $\delta_4$ \\ \hline\hline
$\moda$ $\left[ \text{GeV}^{-3} \right]$         & $15.60(23)$ & $15.16(22)$   & $15.23(22)$   & $15.07(22)$ \\
$\Re(b) \left[\text{GeV}^{-2} \right]$           & $0.690(19)$ & $0.810(17)$   & $0.788(17)$   & $0.819(16)$  \\
$\Im(b) \left[\text{GeV}^{-2} \right]$           & $0.312(30)$ & $0.570(40)$   & $0.545(38)$   & $0.594(40)$  \\
$\Re(A) \left[\text{GeV}^{-3} \right]$           & $0.1251(98)$& $0.1208(97)$  & $0.1208(97)$  & $0.1212(97)$  \\
$\Im(A) \left[\text{GeV}^{-3} \right]$           & $0.025(12)$ & $0.053(12)$   & $0.050(12)$   & $0.056(12)$  \\ \hline
$\left\lvert f_{\phi \pi^0}(0) \right\rvert$ $\left[ \text{GeV}^{-1} \right]$         
                                                 & $0.1351(26)$& $0.1351(26)$  & $0.1351(26)$  & $0.1351(26)$ \\
$\theta_{\phi \pi^0}(0)$                                          & $0.40(17)$  & $0.50(17)$    & $0.48(17)$ & $0.50(17)$  \\ \hline
$N_{ev}\left[10^{6}\right]$         & $6.60(10)$  & $6.60(10)$  & $6.60(10)$ & $6.60(10)$\\
$\chi^2/\text{d.o.f.}$               & $1.01$ & $1.03$ & $1.03$ & $1.03$\\
 \hline
\end{tabular}
\caption{Values of the fitted parameters (upper part) and of the different $\chi^2$ functions (lower part) for the four different fits considered in this work. The error represents our $1\sigma$ statistical uncertainty, and is computed through MC resampling. 
\label{table:parameters}}
\end{table}

We perform four separate fits to the data, each using one of the parameterizations of the phase-shift as input. The resulting values for the fit parameters are shown in Table \ref{table:parameters}. Additionally, we perform an uncertainty estimation using bootstrap with Monte Carlo (MC) resampling \cite{Efron:1986hys,JPAC:2021rxu}, comprising $10^5$ pseudo-data sets for each fit. As a consistency check, the errors provided in Table \ref{table:parameters} through this method are indeed very similar to the parabolic errors and those obtained with \texttt{Minos}. As illustrated in Table\,\ref{table:parameters}, the data description is very good, with $\chi^2/\text{d.o.f.}$ very close to $1$ in all cases. We note that the values of the parameter $b$ agree with each other for the $\delta_{2,3,4}$ parametrizations, although these values differ with that obtained with the $\delta_{1}$ parametrization. This is reasonable, since the range of validity of those parameterizations is quite different, and hence the parameter $b$ might be reabsorbing those differences. The additional parameter we included, representing the total number of events $N_{\text{ev}}$, agrees with the number of events computed from the experimental bins that we are actually using: $6.664(10) \cdot 10^{6}$ events.

\begin{figure}\centering
\includegraphics[height=6cm,keepaspectratio]{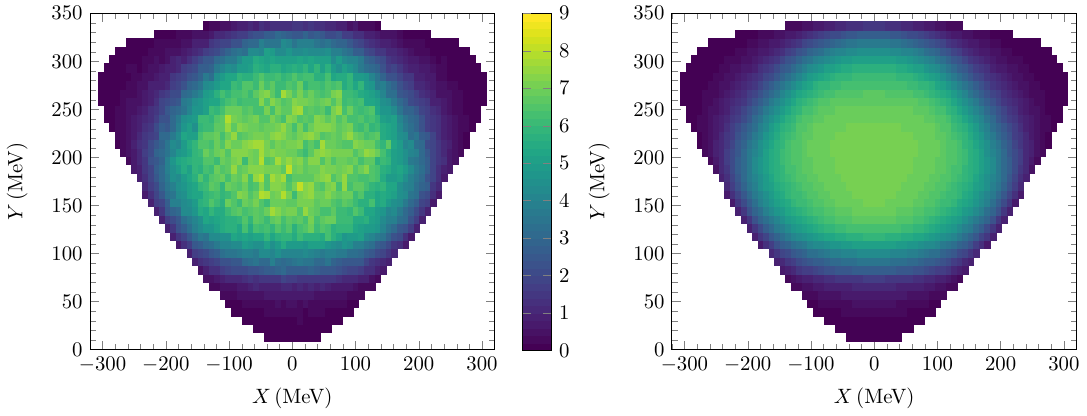}\\
\includegraphics[height=6cm,keepaspectratio]{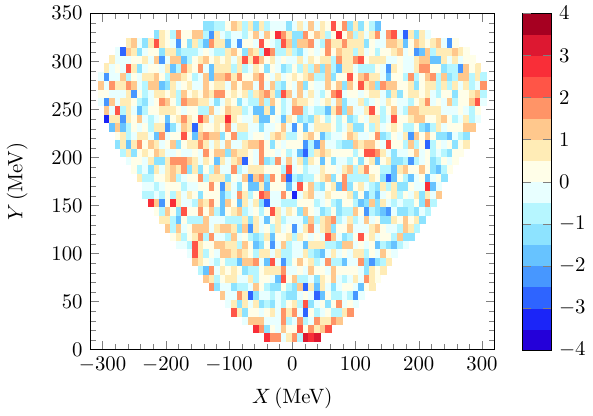}
\caption{Dalitz plot measured by the KLOE experiment \cite{KLOE:2003kas} (upper left) and the same plot using our fitted values (upper right) in terms of the $X,Y$ variables. Furthermore, we show here the values of our statistical residuals (lower center) in the same phase-space region. 
\label{fig:DalitzPlot}}
\end{figure}

In Fig.\,\ref{fig:DalitzPlot}, we present the $\phi \to 3\pi$ DP. The upper panels show the experimental (left) and theoretical (right) distributions obtained in arbitrary units. The theoretical distribution is derived using the phase shift $\delta_{1}(s)$ as input; however, the results are quite similar across all parameterizations used. A visual inspection indicates their similarity. A quantitative comparison between the two is illustrated in the lower center plot, which presents the residuals---defined as the difference for each bin between the theoretical and experimental values divided by the experimental error. As we observe, the residuals do not show any patterns, suggesting that no regions in the DP are described better than others. Furthermore, we have verified that the distribution of residuals approximately follows a normal distribution centered around zero, with a standard deviation of one. In summary, all of these observations, combined with the values of $\chi^2/\text{d.o.f.} \simeq 1$, confirm that the description of the event distribution throughout the DP region is very accurate.

\begin{figure}\centering
\includegraphics[height=4.9cm,keepaspectratio]{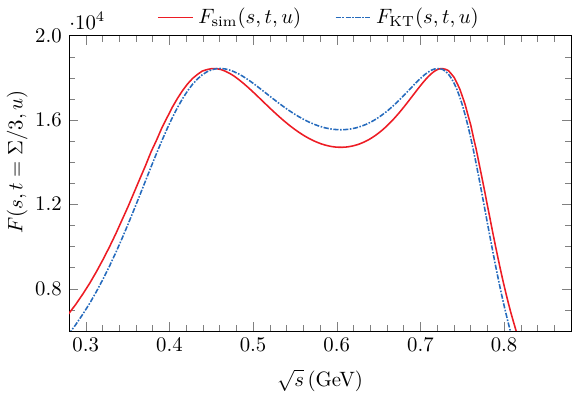}%
\hspace{0.25cm}%
\includegraphics[height=4.9cm,keepaspectratio]{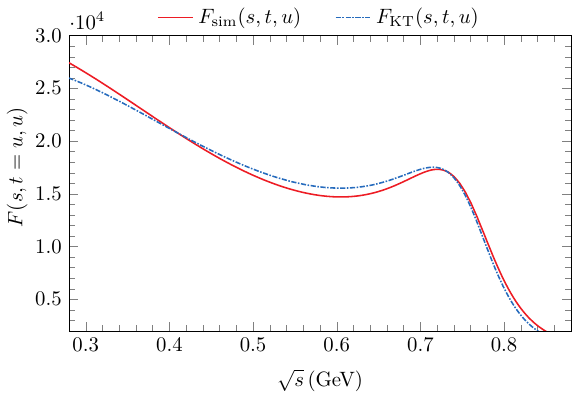}\\%
\includegraphics[height=5.5cm,keepaspectratio]{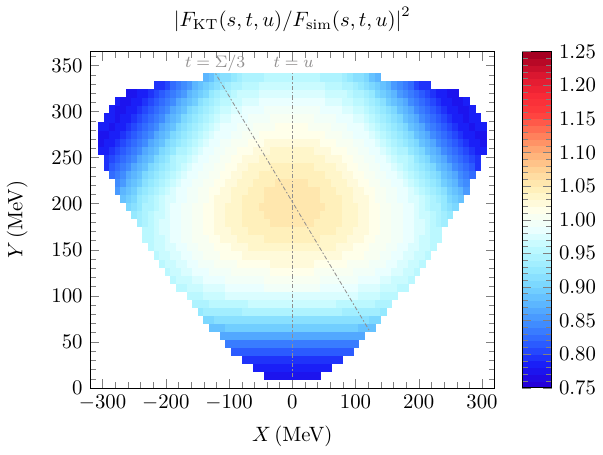}%
\caption{Comparison of the amplitude $F(s,t,u)$ computed from the simplified model in Eq.\,\eqref{eq:Fsimplified} ($F_\text{sim}$) and from our full KT calculation ($F_\text{KT}$). In the top panels we show $\left\lvert F_\text{sim} \right\rvert^2$ (red solid) and $\left\lvert F_\text{KT} \right\rvert^2$ (blue dashed) across two particular lines of the DP, namely $t=\Sigma/3$ (left) and $t=u$ (right), with $\Sigma$ defined in Eq.\,\eqref{eq:sum_mandelstam}. These two lines are also represented inside the DP region in the bottom panel, where the modulus squared of the ratio is represented.%
\label{fig:DPFsim}%
}
\end{figure}

To emphasize the importance of the inclusion of crossed-channel effects through KT equations, one could attempt to describe the $\phi \to 3\pi$ DP with the following simplified amplitude:
\begin{equation}\label{eq:Fsimplified}
F_\text{sim}(s,t,u)=a' \left( \Omega(s) + \Omega(t) + \Omega(u) \right)\,,
\end{equation}
where $a'$ would be a constant to be fixed from the $\phi \to 3\pi$ partial decay width. However, being $a'$ a global constant, its value does not affect the DP shape. This simplified amplitude satisfies crossing symmetry and takes into account $\pi\pi$ scattering in the $I=J=1$ wave but does not satisfy the proper unitarity relation, \textit{i.e.} Eq.\,\eqref{eq:Disc}.

In Fig.\,\ref{fig:DPFsim} we compare $F_\text{sim}(s,t,u)$ with $F(s,t,u)$ computed with KT equations, which we denote $F_\text{KT}$. In the upper panels we show the variation with $\sqrt{s}$ of $F(s,t,u)$ computed accross two particular lines of the Dalitz plot, namely $t=\Sigma/3$ (left) and $t=u$ (right). Both curves show similarities, since the main contribution is given by the Omnès function $\Omega(s)$, dominated by the $\rho$ resonance at around $\sqrt{s} \simeq 0.75\,\text{GeV}$. However, numerical differences can be also appreciated. For instance, around the center $\sqrt{s} \simeq 0.6\,\text{GeV}$, the difference is around $5\%$, and it is even larger at the edges of the plots. In the bottom panel we show the ratio $\left\lvert F_\text{sim}(s,t,u)/F_\text{KT}(s,t,u) \right\rvert^2$ across the full Dalitz plot region. As can be seen, the ratio is larger than one in the center and smaller in the edges, and the difference can reach up to $25\%$. Therefore, the number of events at each bin in the Dalitz plot computed with both parameterizations would be significantly different. We recall here that, as we have discussed above (see Fig.\,\ref{fig:DalitzPlot} and its discussion), our full calculation with $F_\text{KT}$ agrees very well with the data.

This comparison is furthermore relevant in relation with previous works about $\omega \to 3\pi$. Experimentally, the DP parameters of this decay seemed to be in better agreement with the calculations obtained without KT equations, \textit{i.e.}, neglecting altogether crossed-channel effects and using the amplitude in Eq.\,\eqref{eq:Fsimplified}, than with those obtained with UDR KT equations \cite{Niecknig:2012sj,Danilkin:2014cra,WASA-at-COSY:2016hfo,BESIII:2018yvu} (see Ref.\,\cite{JPAC:2020umo} for further details). In Ref.\,\cite{JPAC:2020umo} it is shown that this is only true for the case of UDR, and the agreement is restored when ODR are used. Furthermore, it is also shown that the coincidence of the results with ODR KT equations and the model of Eq.\,\eqref{eq:Fsimplified} is due to coincidental cancellations across the Dalitz plot in the KT equations. More specifically, for the $\omega \to 3\pi$ case, there is much more similarity between $F_\text{sim}(s,t,u)$ and $F_\text{KT}(s,t,u)$ than in the $\phi \to 3\pi$ case, as can be appreciated by comparing Fig.\,10 in Ref.\,\cite{JPAC:2020umo} and the upper panels of Fig.\,\ref{fig:DPFsim} in our work. This demonstrates that the aforementioned cancellations are not general, but particular to the case of the $\omega \to 3\pi$. In summary, the comparison in Fig.\,\ref{fig:DPFsim} clearly calls for a proper inclusion of the crossed-channel effects through KT equations, as we have done in this work.

\begin{figure}\centering
\includegraphics[height=9cm,keepaspectratio]{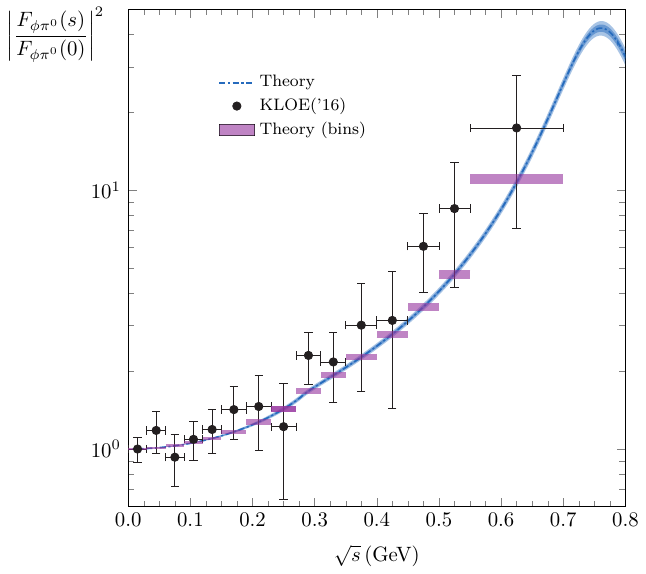}
\caption{Normalized transition form factor in logarithmic scale as a function of $\sqrt{s}$ of the process $\phi \to \pi^0\gamma^\ast$. The black points represent the experimental data by KLOE \cite{KLOE-2:2016pnx}, whereas the purple rectangles represent the results of our fit, with the parameters given in Table\,\ref{table:parameters} ($\delta_1(s)$ in this case) and computed as an average over each bin [\textit{cf.} Eq.\,\eqref{eq:TFF_average}]. The blue dashed line represents the same TFF but computed directly as a continuous function of energy [\textit{cf.} Eq.\,\eqref{eq:Transition}]. The darker, inner band shows the $1\sigma$ statistical uncertainty, whereas the lighter, outer band represents the maximum difference found when using the different parameterizations of the phase shifts.
\label{fig:TFF1GeV}}
\end{figure}

\begin{figure}\centering
\includegraphics[height=9cm,keepaspectratio]{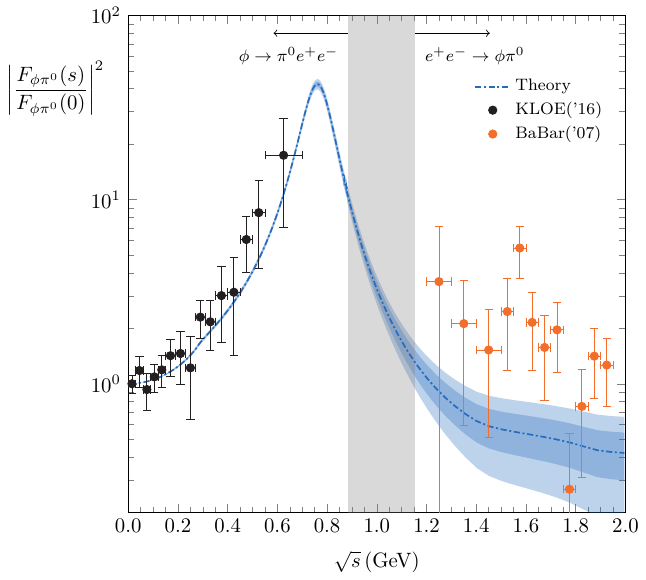}
\caption{Normalized transition form factor modulus squared in logarithmic scale as a function of $\sqrt{s}$ of the process $\phi \to \pi^0\gamma^\ast$. The details are as in Fig.\,\ref{fig:TFF1GeV}, except that we display here the decay ($\phi \to \pi^0 e^+ e^-$) and scattering ($e^+ e^- \to \phi \pi^0$) regions. In addition to the KLOE data (black points, Ref.\,\cite{KLOE-2:2016pnx}), we also show the BaBar data (orange points, Ref.\,\cite{BaBar:2007ceh}) in the scattering region. The latter are not included in the fit, and are only shown here as a reference (see the text for details).
\label{fig:TFF2GeV}}
\end{figure}

In Fig.\,\ref{fig:TFF1GeV}, we show the absolute value squared of the (normalized) TFF resulting from our fits. As seen, our description shows a nice agreement with the experimental data from KLOE\,\cite{KLOE-2:2016pnx}, although falls slightly below the central values. Our results are similar to those in Refs.\,\cite{Schneider:2012ez,Danilkin:2014cra}, which are also shown in Fig.\,4 of Ref.\,\cite{KLOE-2:2016pnx}. In Fig.\,\ref{fig:TFF2GeV}, we show the same functions but in a larger range of $0<\sqrt{s}<2\,\text{GeV}$. We also include in this plot the data provided by BaBar\,\cite{BaBar:2007ceh}, although we have not included them in our fits, since they are probably beyond the limit of applicability of our formalism. We can see that our results follow a reasonable trend in the scattering region, despite the fact that the KT equations are intended only for a low-energy regime.

The inclusion of the $\rho^0$--$\omega$ mixing through Eq.\,\eqref{eq:resomega} is not visible in the DP (Fig.\,\ref{fig:DalitzPlot}) nor in the TFF (Figs.\,\ref{fig:TFF1GeV} and \ref{fig:TFF2GeV}). However, as highlighted in Ref.\,\cite{Niecknig:2012sj}, it is necessary for a proper description of the DP data, because the $\omega$ meson is quite narrow ($\Gamma_\omega = 8.68(13)\,\text{MeV}$) and hence the effect is concentrated in a band $\sqrt{s} \simeq M_\omega=782.66(13)\,\text{MeV}$. In the variable $Y$ [\textit{cf.} Eq.\,\eqref{eq:Ydefinition}], this corresponds to the value $Y \simeq 79\,\text{MeV}$, and thus it affects the two horizontal slices of bins immediately below and above this $Y$ value. Without the inclusion of this effect, the $\chi^2$ gets significantly worse. As said, it is not visible either in the TFF, where only the $\rho$-peak can be seen, but it is also true that there are no data in the region, which is, ultimately, what enables one to follow the minimal approach in Eq.\,\eqref{eq:resomega}. We note that the inclusion of the mixing through Eq.\,\eqref{eq:resomega} is done \textit{after} the iterative calculation of KT equations. We have checked that, if the effect is included in the iterative procedure, no significant changes are observed. Future measurements of this reaction in the $\rho^0$--$\omega$ region will allow to pursue more sophisticated ways to treat the mixing, see \textit{e.g.} Ref.\,\cite{Dias:2024zfh} and references therein.

\begin{table}\centering
\begin{tabular}{|r|l|l|l|l|} \hline
 &                                                         $\delta_1$  & $\delta_2$ & $\delta_3$ & $\delta_4$ \\ \hline\hline
$\Re(b_{\text{sum}}) $           & $0.613$ & $0.658$   & $0.626$   & $0.621$  \\
$\Im(b_{\text{sum}}) $           & $0.506$ & $0.525$   & $0.523$   & $0.524$  \\ \hline

$\Re(b) $           & $0.690(19)$ & $0.810(17)$   & $0.788(17)$   & $0.819(16)$  \\
$\Im(b) $           & $0.312(30)$ & $0.570(40)$   & $0.545(38)$   & $0.594(40)$  \\

 \hline
\end{tabular}
\caption{Values of the parameter $b$ (in $\text{GeV}^{-2}$ units) given by the sum rule (upper part) and obtained through the global fit (lower part) for each phase shift parameterization $\delta_i$. The lower part corresponds to the values in Table~\ref{table:parameters}.
\label{table:paramSUM}}
\end{table}

We now discuss the value of the subtraction constant $b$. We present the fitted value in the lower part of Table \ref{table:paramSUM} (taken from Table \ref{table:parameters}), while the upper part displays the value it takes when the isobar $F(s)$ satisfies a UDR, \textit{i.e.}, the sum rule value $b_\text{sum}$ in Eq.\,\eqref{eq:sumrule}. Note that $b_\text{sum}$ depends on the phase shift used as input, and thus a value is shown for each of the parameterizations $\delta_i$. We observe that, for $\delta_1$, the real parts of $b$ and $b_\text{sum}$ are closer than the imaginary parts, while the opposite holds for parameterizations $\delta_{2,3,4}$. We note that the fitted values of $b$ differ sufficiently from $b_\text{sum}$, which justifies the convenience of introducing a subtraction in KT equations,  \textit{i.e.},  using ODR instead of UDR.

We now compare our study of $\phi \to 3\pi,\,\pi^0\gamma^\ast$ to the corresponding analysis of the $\omega \to 3\pi,\,\pi^0\gamma^\ast$ performed in Ref.\,\cite{JPAC:2020umo}. We observe that the difference between $b$ and $b_\text{sum}$ in the $\phi \to 3\pi$ analysis in this work is considerably milder than in the $\omega \to 3\pi$ case. In Ref.~\cite{JPAC:2020umo}, the value obtained for the sum rule is $b_{\text{sum}} = 0.544 + 0.082i$, while the fitted value is $b_\text{fit} = -0.341(40) + 2.62(79)i$. As discussed in Ref.~\cite{JPAC:2020umo}, employing ODR for KT equations is essential in that instance to achieve a satisfactory description of the $\omega \to 3\pi$ DP parameters determined by BESIII \cite{BESIII:2018yvu}.\footnote{The experimental measurement of WASA-at-COSY \cite{WASA-at-COSY:2016hfo} of the same DP parameters was not utilized in the analysis.} Once ODR are employed, the fits in Ref.~\cite{JPAC:2020umo} are primarily influenced by the high-energy data of the TFF at the highest energies in the decay ($\omega \to \pi^0 \gamma^\ast \to \pi^0 e^+e^-$) region, around $0.6$--$0.7\,\GeV$, particularly from the NA60 collaboration \cite{NA60:2016nad}, which presents much smaller uncertainties compared to the data from MAMI \cite{Adlarson:2016hpp}. Indeed, those higher points lie outside the uncertainty band of the fit (see Fig.\,7 of Ref.\,\cite{JPAC:2020umo}). In contrast, our analysis for the $\phi$ TFF in Fig.\,\ref{fig:TFF1GeV} is consistent with all data points. Furthermore, the high-energy trends of the $\omega$ and $\phi$ TFF are also different (compare Fig.\,11 of Ref.\,\cite{JPAC:2020umo} with Fig.\,\ref{fig:TFF2GeV} from our manuscript, respectively): while the $\phi$ TFF has a reasonable extrapolation to higher energies, the $\omega$ TFF of Ref.\,\cite{JPAC:2020umo} appears to be significantly above the data in the scattering region, as measured by CMD \cite{CMD-2:2003bgh} and SND \cite{Achasov:2013btb}. These striking contrasts between the $\omega$ and $\phi$ cases are quite intriguing, especially since one might expect their behaviors to be similar, and it would be beneficial if these discrepancies spurred new experimental analyses.

In this regard, it is worth recalling here the case mentioned in the Introduction of the $J/\psi \to 3\pi$ and $J/\psi \to \pi^0 \gamma^\ast$, studied similarly in Ref.\,\cite{JPAC:2023nhq}, and where DP data from BESIII collaboration \cite{BESIII:2012vmy} were fitted. Interestingly, the recent BESIII data for $J/\psi \to \pi^0 \gamma^\ast$ \cite{BESIII:2025xjh} are in good agreement with the predictions of Ref.\,\cite{JPAC:2023nhq}. This might point out the fact that the $J/\psi$ analysis is closer to the $\phi$ case presented here than to the $\omega$ one.

\section{Summary and Outlook}\label{sec:summary}

In this work, we analyze the $\phi \to 3\pi$ decay and the $\phi \to \pi^0 \gamma^\ast$ transition form factor using the Khuri-Treiman formalism. This consistent framework incorporates crossing and maintains analyticity and unitarity. The model successfully reproduces the KLOE data for the Dalitz plot and the transition form factor, utilizing once-subtracted dispersion relations instead of unsubtracted ones. Notably, while our analysis of the transition form factor focuses on the decay region, the model also captures the trend in the scattering regime ($e^+ e^- \to \phi \pi^0$), suggesting its potential for broader applications. These results reinforce the applicability of the framework to three-body decays. A comparison with the analogous $\omega \to 3\pi$ case, previously analyzed with the same framework, reveals a striking difference, particularly in the value of the subtraction constant introduced in the Khuri-Treiman equations. In the $\phi$ case, the fitted value is close to the one determined by the sum rule [\textit{cf.} Eq.\,\eqref{eq:sumrule}], which would reduce the once-subtracted dispersion relation to an unsubtracted form. This sharply contrasts with the $\omega$ case, where the fitted value significantly deviates from the sum rule value. As discussed in Ref.\,\cite{JPAC:2020umo}, this deviation is primarily driven by the transition form factor high-energy data points from the NA60 collaboration. Additional high-precision data, especially for the $\omega$ decay, will be crucial for refining the theoretical framework and enhancing our understanding of these processes. These intriguing theoretical differences should encourage and stimulate new experimental analyses.

\acknowledgments

This work is supported by the Spanish Ministerio de Ciencia e Innovaci\'on (MICINN) under contracts PID2020-112777GB-I00, PID2023-147458NB-C21 and CEX2023-001292-S; by Generalitat Valenciana under contracts PROMETEO/2020/023 and  CIPROM/2023/59. M.\,A. acknowledges financial support through GenT program by Generalitat Valencia (GVA) Grant No.\,CIDEGENT/2020/002, Ramón y Cajal program by MICINN Grant No.\,RYC2022-038524-I, and Atracción de Talento program by CSIC PIE 20245AT019. %
The work of S.~G-S., N.H and V.M. is supported by MICIU/AEI/10.13039/501100011033 and by FEDER UE through grants PID2023-147112NB-C21; and through the ``Unit of Excellence Mar\'ia de Maeztu 2020-2023'' award to the Institute of Cosmos Sciences, grant CEX2019-000918-M. Additional support is provided by the Generalitat de Catalunya (AGAUR) through grant 2021SGR01095. S.~G-S.~and V.~M. are Serra H\'{u}nter Fellows. V.~M. acknowledges support from Spanish national Grant CNS2022-136085.
G.M. and A.P.S. were supported by the U.S. Department of Energy contract \mbox{DE-AC05-06OR23177}, under which Jefferson Science Associates, LLC operates Jefferson Lab. A.P.S. acknowledges support from \mbox{DE-FG02-87ER40365}. This work contributes to the aims of the U.S. Department of Energy ExoHad Topical Collaboration, contract DE-SC0023598.%

\bibliographystyle{apsrev4-1_MOD}
\bibliography{refs}

\end{document}